\documentclass[prl,superscriptaddress,twocolumn,nofootinbib,showpacs]{revtex4}

\usepackage[latin1]{inputenc}
\usepackage{graphicx,psfrag}
\usepackage{bm,longtable}

\usepackage{txfonts}

\newcommand{\E}{\mathrm{e}}

\newcommand{\average}[1]{\left\langle{#1}\right\rangle}

\newcommand{\eij}{\epsilon_{ij}}
\newcommand{\dij}{\Delta_{ij}}

\newcommand{\lij}{l_{ij}}
\newcommand{\tT}{\widetilde T}

\newcommand{\tu}{\tau_{\mathrm u}}
\newcommand{\p}[1]{\left({#1}\right)}
\newcommand{\pq}[1]{\left[{#1}\right]}
\newcommand{\pg}[1]{\left\{{#1}\right\}}

\newcommand{\Eu}{E_{\mathrm u}}
\newcommand{\xu}{x_{\mathrm u}}
\newcommand{\Lu}{L_{\mathrm u}}

\begin{document}
\title{An Ising-like model for protein mechanical unfolding}
\author{A. Imparato}
\affiliation{Dipartimento di Fisica and CNISM, Politecnico di Torino,
  c. Duca degli Abruzzi 24, Torino, Italy}
\affiliation{INFN, Sezione di Torino, Torino, Italy}
\author{A. Pelizzola}
\affiliation{Dipartimento di Fisica and CNISM, Politecnico di Torino,
  c. Duca degli Abruzzi 24, Torino, Italy}
\affiliation{INFN, Sezione di Torino, Torino, Italy}
\author{M. Zamparo}
\affiliation{Dipartimento di Fisica and CNISM, Politecnico di Torino,
  c. Duca degli Abruzzi 24, Torino, Italy}

\begin{abstract}
The mechanical unfolding of proteins is investigated by extending  
the Wako-Sait\^o-Mu\~noz-Eaton model, a simplified protein model
with binary degrees of freedom, which has proved successful in describing
the kinetics of protein folding. Such a model is generalized by including the effect of an 
external force, and its thermodynamics turns out to be exactly solvable. 
We consider two molecules, the 27th immunoglobulin domain of titin
and protein PIN1. In the case of titin we determine equilibrium
force-extension curves and study nonequilibrium phenomena in the
frameworks of dynamic loading and force clamp protocols, verifying
theoretical laws and finding the position of the kinetic barrier which
hinders the unfolding of the molecule. The PIN1 molecule is used to 
check the
possibility of computing the free energy landscape as a function of
the molecule length by means of an extended form of the Jarzynski equality.
\end{abstract}
\maketitle

Manipulation experiments on single biomolecules have greatly increased
our knowledge of the structural properties of such molecules. In a
typical experiment a controlled force is applied to one of the free
ends of the molecule, and the induced elongation is measured.  Such
experimental techniques have been used to probe the structure of
proteins 
\cite{rgo,png}
and nucleic acids \cite{bus1,bus2}.
According to the common interpretation the unfolding of a molecule being pulled from one
of its ends is hindered by kinetic barriers associated with the
strongest linkages which serve to stabilize the molecular structure.
The breaking of a
molecular bond can thus be viewed as the overcoming of a kinetic
barrier. 
It has been
argued \cite{ER} that the study of the kinetics of bond breaking under
different loading rates can provide much information about the 
internal structure of molecules, and in particular allows one to measure the
strength of the molecular bonds, and to associate to them a position
along the molecular structure.
In the case of simple molecules, such as RNA hairpins, it is easy to 
obtain information on the molecular structure by pulling experiments \cite{bus1}. However,
when one deals with large molecular structures, such as multi-domain proteins, the inference of structural characteristics from
the unfolding kinetics can prove a difficult task.
Therefore, the study of simple models for the unfolding of proteins, whose microscopic native structures are known {\it a priori},
is highly desirable: investigating the kinetics of such models can shed new light on the relations between the experimentally observed unfolding features and the molecular structures. 

The experiments discussed above  are usually performed in non-equilibrium conditions: because
of technical limitations the pulling process is faster than the
typical molecular relaxation time. 
The problem of irreversibility of unfolding processes can be avoided
by using the remarkable equality introduced by Jarzynski \cite{jarz},
which allows one to measure the free energy difference between the folded
and the unfolded state of a biomolecule \cite{jarzexp}. By using 
an extended form of the Jarzynski equality (JE)  the free energy landscape of simple models of biopolymers has been probed
as a function of the molecular elongation 
\cite{HS,seif, alb1}.
Although this approach appears very promising, it still  has to be tested on systems of increasing complexity.

Here we approach the mechanical unfolding problem by means of a
suitable generalization of the Wako-Sait\^o-Mu\~noz-Eaton (WSME)
protein folding model \cite{WS1,ME1,ME3}. This is a simplified
statistical mechanical model where a binary variable is associated to
each peptide bond. 
The equilibrium thermodynamics has been
solved exactly \cite{Ap1,Ap2} and the model has been quite successful
in describing the kinetics of protein folding
\cite{Amos,ItohSasai1,Ap3,CCBM}
and has also found applications in different fields
(see \cite{TD3} and references therein).

In the present paper we first extend the WSME model by considering the effect
of an external force, and show that the equilibrium properties of this new model
can be computed exactly, similarly to the original WSME model.
In order to mimic the mechanical unfolding of proteins, we use computer simulations and study the unfolding kinetics of our model, both in the cases of constant force and dynamic loading.
Finally we probe the free energy landscape of a model protein, exploiting the JE.

The WSME model describes a protein of $N+1$ aminoacids as a chain of
$N$ peptide bonds (connecting consecutive aminoacids) that can live in
two states (native and unfolded) and can interact only if they are in
contact in the native structure and if all bonds in the chain between
them are native. To each bond is associated a binary variable $m_{k}$,
with values $0,1$ for unfolded and native state respectively. The
effective Hamiltonian of the model reads
\begin{equation}
\mathcal H_0(\{m_k\})=\sum_{i=1}^{N-1} \sum_{j=i+1}^N \eij \dij
\prod_{k=i}^j m_k -k_B T \sum_{i=1}^N q_i(1-m_i),
\label{H0}
\end{equation} 
where  $k_B$ is the Boltzmann constant, and $T$ is the absolute temperature. 
 The first term assigns an energy $\eij<0$
to the contact (defined as in \cite{ME3,Ap1}) between bonds $i$ and
$j$ if this takes place in the native structure ($\dij=1$ in this case
and $\dij=0$ otherwise). The second term represents the entropic cost
$q_{i}>0$ of ordering bond $i$. In Ref.~\cite{Ap1}  it is shown how to
compute exactly the partition function $Z = \sum_{\{m_k\}}
\exp[ - \beta \mathcal H_0(\{m_k\})]$ and the corresponding thermal
averages. Here and in the following, the quantity $\beta$ indicates the inverse temperature $\beta=1/(k_B T)$. 

In order to couple the protein to an external force
 we assume that  a
configuration $\{m_k\}$ of the model defines a sequence of native
stretches, separated by unfolded peptide bonds (see inset of Fig.~\ref{eq_1TIT}). Peptide bonds $i$ and $j$ delimit a native stretch if and
only if $m_i = m_j = 0$ and $m_k = 1$ for $i < k < j$.  A native
stretch is regarded as a rigid portion of the molecule, even under
application of the external force, with an end-to-end length $\lij$.
If $j = i+1$
the stretch reduces to a single aminoacid, which is also regarded as a rigid structure, with length $l_{i,i+i}$.
The values of the parameters $\lij$ are taken from the native structure of the protein \cite{append}. 
Boundary conditions are
introduced through the dummy bonds $m_0 = m_{N+1} = 0$.
Given the direction of the external force, we assume
that a native stretch,  or a single aminoacid delimited by two successive unfolded bonds, can only take two orientations, parallel or
antiparallel to the force, so that they contribute $\pm \lij$ to the
length of the molecule.  Therefore, given a
configuration $\{m_k\}$ of the peptide bonds, we introduce a variable
$\sigma_{ij}$ for each rigid portion of the molecule, either a native stretch or an aminoacid delimited by two successive unfolded bonds, taking values $+1$ (rigid portion 
parallel to the force) or $-1$ (antiparallel). Thus the end-to-end
length of the molecule, in the force direction, reads
\begin{equation}
L(\{m_k\}, \{ \sigma_{ij}\}) = \sum_{0 \le i < j \le N+1} \lij
\sigma_{ij} (1-m_i) (1-m_j) \prod_{k=i+1}^{j-1} m_k.
\label{L_def}
\end{equation}
Let us define  the Hamiltonian $\mathcal H$ as the sum of the interaction energy term, contained in the effective Hamiltonian 
$\mathcal H_0$ (\ref{H0}), and of the term $-f L$, which  takes into
account the effect of the external force
$\mathcal H(\{m_k\},\{\sigma_{ij}\},f)= \sum_{i<j}
\eij \dij \prod_{k=i}^j m_k - f L(\{m_k\}, \{\sigma_{ij}\}).$
Note that the definition of molecular length (\ref{L_def}) is such that the set of variables $\{\sigma_{ij}\}$ is dynamically defined by
the bond configuration $\{m_k\}$: for each configuration $\{m_k\}$, we consider only those variables $\sigma_{ij}$ such that 
$(1-m_i)(1-m_j)m_{i+1}m_{i+2}\dots m_{j-1}=1$.
One can easily sum over the variables $\{\sigma_{ij}\}$:
$\sum_{\{\sigma_{ij}\}} \exp[- \beta \mathcal
H(\pg{m_k},\{\sigma_{ij}\},f)] = \exp[- \beta \mathcal H_{\rm eff}(\pg{m_k},f)]$,
where the new effective Hamiltonian $\mathcal H_{\rm eff}$ is given by
\begin{eqnarray}
&& \mathcal H_{\rm eff}(\pg{m_k},f) = \sum_{i=1}^{N-1} \sum_{j=i+1}^N
\eij \dij \prod_{k=i}^j m_k\nonumber \\
&& -  k_BT \sum_{i < j } 
\ln \left[ 2 \, {\rm cosh} \left( \beta f l_{ij} \right) \right]
(1-m_i) (1-m_j) \prod_{k=i+1}^{j-1} m_k,
\label{heff}
\end{eqnarray}
and, as a function of $\{m_k\}$, has the same structure as  the WSME model (see Eq.~(\ref{H0})).  Therefore  its
equilibrium thermodynamics  is exactly solvable, as discussed in Ref.~\cite{Ap1}. 
In the case $f = 0$, $\mathcal H_{\rm eff}$ reduces to Eq.~(\ref{H0}) (up to an
additive constant), with $q_i=\ln 2$.
Thus, the effective
entropic terms can be viewed as resulting from microscopic
orientational degrees of freedom of the native stretches. 

In the following, the quantity $\epsilon$ will indicate the system energy scale.
This quantity, together with 
the interaction energies $\eij$,  will be 
chosen as in \cite{ME3,Ap1,Ap3}, see also \cite{append}.  We also introduce the reduced temperature in terms of the energy scale: $\tT= k_B T/\epsilon$ . 
The time scale will be indicated by $t_0$. 

Following Ref.~\cite{Ap1}, for any choice of the model parameters, one can
calculate the equilibrium value of the order parameter $m =
1/N\sum_k \langle m_k \rangle$ and of the molecule length, as defined by Eq.~(\ref{L_def}), for
varying force and temperature. We first consider the 
titin immunoglobulin domain I27 (89 aminoacids, pdb code 1TIT),   which has been widely studied
both experimentally \cite{rgo} and theoretically \cite{teotitin}.
The energy scale for such a molecule is taken to be $\epsilon/k_B =43.1$~K, while the melting temperature is $\tT_m\simeq8.03$ \cite{append}. In figure \ref{eq_1TIT}, the root mean square length of the 1TIT molecule is
plotted as a function of the external force $f$ for different temperatures.
\begin{figure}[h]
\center
\psfrag{lij}[cb][cb][1.]{$\lij$}
\psfrag{L1}[tt][tt][1.]{$L$}
\psfrag{f1}[cb][cb][1.]{$f$}
\psfrag{L2}[ct][ct][1.]{$\sqrt{\average{L^2}}\,  (\AA)$}
\psfrag{f}[ct][ct][1.]{$f$ (pN)}
\psfrag{T4}[rc][rc][.8]{$\tT=4$}
\psfrag{T6}[rc][rc][.8]{$\tT=6$}
\psfrag{T8}[rc][rc][.8]{$\tT=8$}
\psfrag{T10}[rc][rc][.8]{$\tT=10$}
\includegraphics[width=8cm]{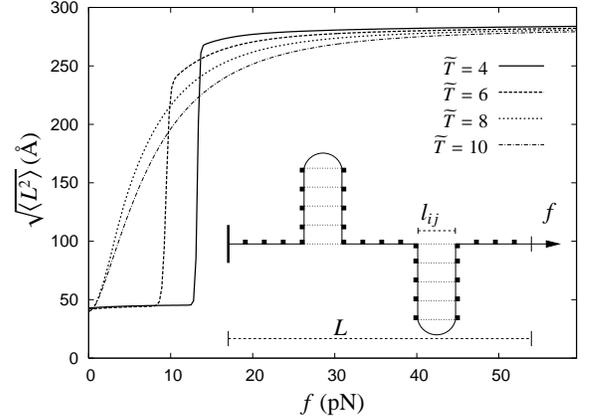}
\caption{
Root mean square length of the 1TIT molecule as a function of the external force $f$, for different temperatures ($L$ is defined by Eq.~(\ref{L_def})). Inset: Cartoon of the model protein under loading.}
\label{eq_1TIT}
\end{figure}
The plateau appearing in Fig.~\ref{eq_1TIT}, in the low temperature regime, corresponds to the overall alignment of the molecule in  the native configuration ($m\simeq1$) to the force direction.
Having introduced the molecular length we have built a framework
within which the mechanically induced protein unfolding can be
simulated by applying an external force.  In the following, two
manipulation schemes will be considered: in the first one the molecule is
manipulated in a ``force-clamp", where a sudden force is applied to one of the
molecule's free ends, in the second one  a time-varying force is applied,
so that the load on the molecule increases gradually. In order to study the molecular unfolding,   we run Monte
Carlo simulations with Metropolis kinetics using the  Hamiltonian 
 $\mathcal H(\pg{m_k},\{\sigma_{ij}\},f(t))$. In the following the time
scale $t_0$ will correspond to a single Monte Carlo step.
In the force clamp manipulation experiments, the
molecule unfolds after a given time $\tu$ which fluctuates between one realization of the unfolding process and the other,  due to the
stochasticity of the unfolding process.  Unfolding can be viewed as an
activated process \cite{ev2}, whose kinetics is dominated by a
characteristic energy barrier $\Delta \Eu$ placed at the value $\xu$
of the reaction coordinate: therefore, it is usually
assumed that the average unfolding time $\average{\tu}$ follows an
Arrhenius law
$\average{\tu}=\omega_0^{-1}\exp\pq{\beta\p{\Delta \Eu-f\xu}}$,
where $\omega_0$ is a characteristic attempt rate depending on the
microscopic features of the system.  

We simulate force clamp manipulations of the  
 1TIT molecule as follows.  Starting
from thermal equilibrium with
$f=0$, at $t=0$ we apply a non-zero force $f$ and thus measure the
unfolding time $\tu$ as the first passage time of the molecule length
across the threshold value $\Lu$, defined as half the molecule
maximal length, $\Lu=140\, \AA$, see figure \ref{eq_1TIT}.
In figure \ref{fc_1TIT} the unfolding time $\tu$ of the 1TIT molecule, averaged over 1000 independent unfolding trajectories,  is plotted as a function of the applied force $f$,
for three values of the temperature $\tT=4,6,8$.  
From a fit of the data shown
in Fig.~(\ref{fc_1TIT}) to the Arrhenius law, 
we find that the unfolding length is
$\xu\simeq3\, \AA$, and is independent of the temperature, as expected (see caption of fig.~(\ref{fc_1TIT}) for the exact values and uncertainties).  
Note that this value is in good agreement with the experimentally measured
value of the titin unfolding length $\xu=2.5\, \AA$, found in Ref.~\cite{rgo}.
\begin{figure}[h]
\center
\psfrag{Logt}[ct][ct][1.]{$\average{\tu}\, (t_0)$}
\psfrag{f}[ct][ct][1.]{$f$ (pN)}
\psfrag{T8}[rc][rc][.8]{$\tT=8$}
\psfrag{T4}[rc][rc][.8]{$\tT=4$}
\psfrag{T6}[rc][rc][.8]{$\tT=6$}
\includegraphics[width=8cm]{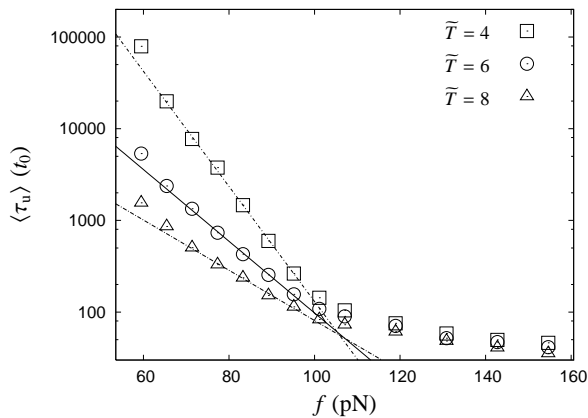}
\caption{Average unfolding time of the 1TIT molecule
as a function of the  force for three values of the temperature. 
The lines are linear fits of the data to
the Arrhenius law discussed in the text. 
From such fits we find the following values of the
unfolding length (in $\AA$): $\tT=4$, $\xu=3.4\pm0.1$; $\tT=6$, $\xu=3.2\pm 0.1$; $\tT=8$,
$\xu=3.0\pm0.1$. 
}
\label{fc_1TIT}
\end{figure}

 

We now consider the case where  
a time-dependent force is
applied to our model molecule and the unfolding time is sampled over 1000 independent trajectories.  
Here the force  increases linearly with time, with a
rate $r$, and
thus the rupture force $f^*$ is given by $f^*=r \tu$, where  unfolding time is 
defined as in the case of the force clamp.
In
refs.~\cite{ev2} it has been argued that, if the energy barrier $\Delta E_u$ is large (compared to the thermal energy $k_B T$) and  rebinding is
negligible, the typical unbinding force of a single molecular bond under dynamic loading is
given by
\begin{equation}
f^*=k_B T/x_u\,  \ln [\beta r  x_u\tau_0]
\label{fstar}
\end{equation} 
where $\tau_0$ is the characteristic unfolding time at zero force, 
$\tau_0=\omega_0^{-1}\exp\pq{\beta \Delta E_u}$.
In Fig.~\ref{dl_1TIT} the breaking force $f^*$ is plotted as a function
of the pulling velocity, for the 1TIT molecule, for three
values of the temperature. 
The value
of the unfolding length obtained by fitting the data to Eq.~(\ref{fstar}) is $x_u\simeq 3\, \AA$, and is
independent of the temperature, as expected (see caption of fig.~\ref{dl_1TIT}
for the exact values). Note that this value agrees with that found with
the force clamp manipulation, and with the experimental value $\xu=2.5\, \AA$ found in Ref.~\cite{rgo}.
\begin{figure}[h]
\center
\psfrag{f}[ct][ct][1.]{$f^*$ (pN)}
\psfrag{v}[ct][ct][1.]{$r$ (pN/$t_0$)}
\psfrag{T6}[cc][cc][.8]{$\tT=6$}
\psfrag{T8}[cc][cc][.8]{$\tT=8$}
\psfrag{T4}[cc][cc][.8]{$\tT=4$}
\includegraphics[width=8cm]{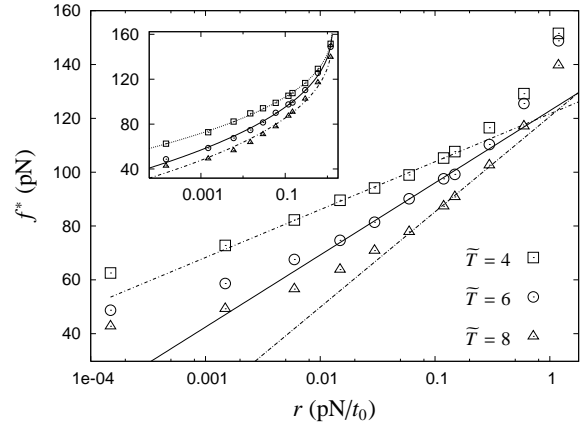}
\caption{Plot of the breaking force  $f^*$ of the 1TIT molecule as a function of the pulling velocity $r$, for the three values of temperature here considered. The lines are  fits to the data in the linear regime defined by Eq.~(\ref{fstar}). From such fits we obtain $x_u=3.1\pm0.1\,  \AA$  for $\tT=4$,  $x_u=3.0\pm0.1\,  \AA$ for $\tT=6$, and $x_u=3.1\pm 0.2\,  \AA$  for $\tT=8$.  
Inset: the lines are fits of the data to the equation defining $\tilde f^*$, see text. 
}
\label{dl_1TIT}
\end{figure}
On the other hand, in recent works \cite{hs_paper}, it has been argued that the rupture force $f^*$ has a more complex expression
$\tilde f^*=\Delta E_u/(\nu x_u)\pg{1-\pq{-k_B T/\Delta E_u \ln\p{\beta r x_u \tau_0 \E^{-\gamma}}}^\nu}$,
where the exponent $\nu$ depends on the microscopic details of the energy landscape, and $\gamma$ is the Euler-Mascheroni constant $\gamma\simeq0.577$.
This equation reduces to Eq.~(\ref{fstar}) in the case $\nu=1$ or in the limit $\Delta E_u \rightarrow \infty$ \cite{hs_paper}.
In the inset of Fig.~\ref{dl_1TIT}, we plot the fits of the rupture force data to the equation defining $\tilde f^*$. Although the agreement of the data with the equation appears to be rather good, the statistical errors of the fit parameters are quite large, since $\tilde f^*$ depends nonlinearly on the set of the unknown parameters. This issue will be addressed in a forthcoming paper.

We now evaluate the effective free energy landscape as a
function of the molecular length $L$ of the extended model here discussed.
Formally, the free energy function $F(L)$ is defined by
$F(L)=-k_B T \ln Z(L)$,
where $Z(L)$ is given by the sum of the Boltzmann weight $\exp[-\beta\mathcal H(\pg{m_k},\{\sigma_{ij}\},f=0)]$, 
over all those configurations $\{m_k\},\{\sigma_{ij}\}$,
whose length 
$L(\{m_k\},\{\sigma_{ij}\})$ (Eq.~(\ref{L_def})) equals the given value $L$.
\begin{figure}[h]
\center
\psfrag{F}[ct][ct][1.]{$F\, (k_B T)$}
\psfrag{L}[ct][ct][1.]{$L\, (\AA)$}
\psfrag{1e2}[cr][cr][.8]{$r=0.6$}
\psfrag{1e3}[cr][cr][.8]{$r=0.06$}
\psfrag{1e4}[cr][cr][.8]{$r=6\cdot 10^{-3}$}
\psfrag{1e5}[cr][cr][.8]{$r=6\cdot 10^{-4}$}
\includegraphics[width=8cm]{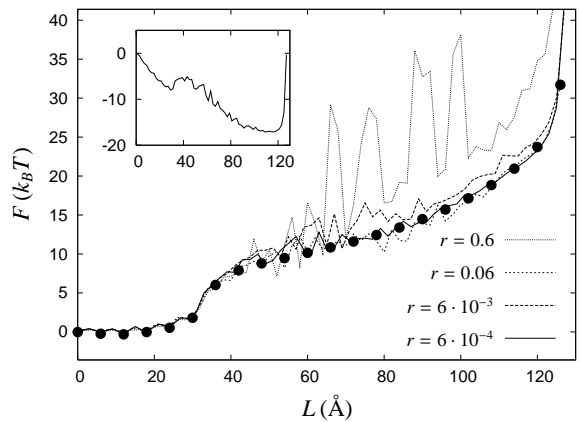}
\caption{Reconstructed free energy landscape $F$ of the PIN1 molecule (lines), as a function of  $L$, for different pulling rates  $r$ (in pN$/t_0$ units) and with $\tT=6$. 
Circles: expected value of $F(L)$ as obtained by direct evaluation of the free energy function.
 Inset: Plot of $F(L)-f L$, for $f=2\, \epsilon/\AA$. The new minimum at $L\simeq120\, \AA$ corresponds to the  length of the molecule in the large force regime (data not shown).
}
\label{land_1I6C}
\end{figure}
It can be shown, that the partition function $Z(L)$ is related to the work done on the molecule during the manipulation via the extended JE \cite{HS} 
\begin{equation}
Z(L)=\average{{\delta(L-L(x_t)) \exp\p{-\beta W_t}}}\exp(-\beta f(t) L)
\label{ext_jarz}
\end{equation} 
where $x_t$ is the system microscopic configuration at time $t$, $L(x)$ is the
macroscopic length corresponding to microscopic configuration $x$,
$W_t$ is the thermodynamic work {\it done} on the system by the external potential,
up to the time $t$, defined by $W_t=\int^t_0 d t'\partial \mathcal H/\partial t'  $, and the average is over all
the trajectories of fixed duration $t$.  
In order to 
 recover the partition function $Z(L)$
from eq.~(\ref{ext_jarz}) we use the procedure introduced and discussed in ref.~\cite{HS}.
The investigation of the free energy landscape of the 1TIT molecule, by using Eq.~(\ref{ext_jarz}), turned out to be a very difficult task.
Indeed, the typical value of the work associated to the  unfolding of this molecule is
of the order of some hundreds of $k_B T$, for the value of $\epsilon$ here used.
Since one has to evaluate $\exp(-\beta W_t)$, in order to exploit
Eq.~(\ref{ext_jarz}), the JE  cannot give a reliable
estimate of the energy landscape of the 1TIT molecule. 
For a discussion of the range of applicability of the JE to microscopic systems see, e.g., \cite{felrev}.
Therefore we consider the PIN1, a smaller  protein whose folding characteristics have already  been studied with the WSME model \cite{Ap3} (pdb code 1I6C, 39 aminoacids, $\epsilon/k_B =44$~K \cite{append}).
Its
free energy landscape as a function of the molecule elongation $L$, as given by Eq.~(\ref{ext_jarz}),  is
plotted in Fig.~\ref{land_1I6C} for different velocities of the pulling
protocol.  It can be seen that the curves $F(L)$ collapse onto the
same curve, as the pulling velocity is decreased. 
This is a clear
signature that the energy landscape is correctly reconstructed, and
its best estimate is the collapse curve, as discussed in Ref.~\cite{alb1}.
On the other hand, the model introduced here is simple enough to allow the exact computation of the function $Z(L)$, and hence
we can obtain the exact value of the function $F(L)$: the agreement with the landscape evaluated by the pulling manipulations is found to be very good, see Fig.~\ref{land_1I6C}.

In conclusion, we have introduced and studied a model of proteins under external loading.
The unfolding length of the titin model is  found to be in good agreement with the experimental one. We believe that this result represents a remarkable validation of the model that we have introduced:
it suggests that our model, although minimal, captures  the basic mechanisms underlying the unfolding process of proteins.
For a small protein, the extended form of the Jarzynski equality
gives an estimate of the  free energy landscape which is 
in good agreement with the expected one.
We believe that our model  can be successfully used to study the interplay between the protein structures
and the kinetics of unfolding and refolding under external loading. As an example, by using computer simulations and applying an external force, one can easily determine which are the main contacts  stabilizing the molecular structures.
Furthermore,  our model is also suitable to study the thermal unfolding of the proteins, in order to compare the thermal and the mechanical unfolding paths.


\newpage

{\Large Appendix to ``An Ising-like model for protein mechanical unfolding''}
\title{Appendix to ``An Ising-like model for protein mechanical unfolding''}
\author{A. Imparato}
\email{alberto.imparato@polito.it}
\affiliation{Dipartimento di Fisica and CNISM, Politecnico di Torino,
  c. Duca degli Abruzzi 24, Torino, Italy}
\affiliation{INFN, Sezione di Torino, Torino, Italy}
\author{A. Pelizzola}
\email{alessandro.pelizzola@polito.it}
\affiliation{Dipartimento di Fisica and CNISM, Politecnico di Torino,
  c. Duca degli Abruzzi 24, Torino, Italy}
\affiliation{INFN, Sezione di Torino, Torino, Italy}
\author{M. Zamparo}
\email{marco.zamparo@polito.it}
\affiliation{Dipartimento di Fisica and CNISM, Politecnico di Torino,
  c. Duca degli Abruzzi 24, Torino, Italy}

\maketitle

In this appendix we briefly review the definition of the parameters $\epsilon$, $\epsilon_{ij}$, $\Delta_{ij}$ in the WSME model, and discuss the definition of
the new parameters $l_{ij}$ which have been introduced in the main text.
\begin{figure*}[h]
\center
\psfrag{lab}[ct][ct][1.]{$l_{i,i+2}$}
\psfrag{lcd}[ct][ct][1.]{$l_{i,i+1}$}
\psfrag{lef}[ct][ct][1.]{$l_{i,j}$}
\psfrag{N}[cb][cb][1.]{$N_i$}
\psfrag{C}[cb][cb][1.]{$C_i$}
\psfrag{Ca}[cb][cb][1.]{$C_{\alpha,i}$}
\psfrag{N1}[cb][cb][1.]{$N_{i+1}$}
\psfrag{C1}[cb][cb][1.]{$C_{i+1}$}
\psfrag{Ca1}[cb][cb][1.]{$C_{\alpha,{i+1}}$}
\psfrag{N2}[cb][cb][1.]{$N_{i+2}$}
\psfrag{C2}[cb][cb][1.]{$C_{i+2}$}
\psfrag{Ca2}[cb][.epscb][1.]{$C_{\alpha,{i+2}}$}
\psfrag{N3}[cb][cb][1.]{$N_{i+3}$}
\psfrag{Ni}[cb][cb][1.]{$N_{j+1}$}
\psfrag{Ci}[cb][cb][1.]{$C_{j+1}$}
\psfrag{Cai}[cb][cb][1.]{$C_{\alpha,{j+1}}$}
\psfrag{Cii}[cb][cb][1.]{$C_{j}$}
\includegraphics[width=16cm]{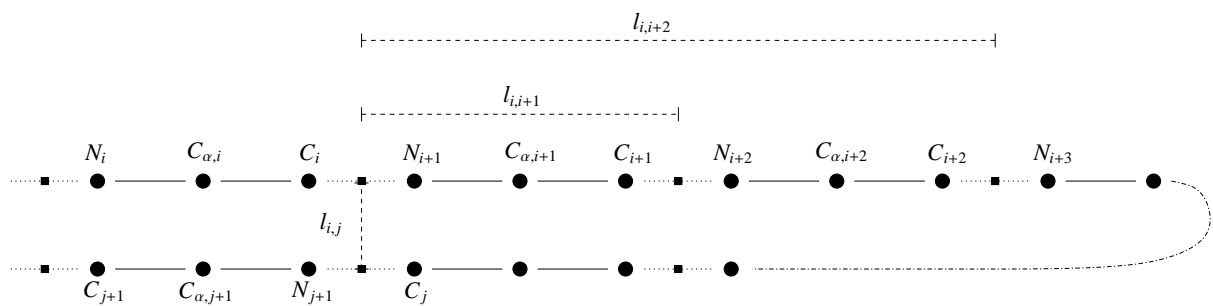}
\caption{Cartoon of the model protein native structure. The lengths $\lij$-s are defined as discussed in the section {\it Model Parameters}.}
\label{lij_fig}
\end{figure*}

The parameters $\Delta_{ij}$ and $\epsilon_{ij}$, appearing in eqs.~(1)-(3) of the main text are chosen following Ref.~\cite{ME3a},
starting from the protein native structure, as given in the Protein Data Bank (pdb in the following, http://www.pdb.org/).
An atomic contact is present ($\Delta_{ij}=1$) if, in the native state of the protein,  at least two atoms from residues $i$ and $j+1$ (with $j+1>i+2$) are closer than $4\, \AA$. In this case $\eij$ is taken to be equal to $k\epsilon$, where $k$
is an integer such that $5(k-1)<n_{at}\leq 5k$, and $n_{at}$ is the number
of atomic contacts. 
As an example, in table \ref{tab_app}, the values of the contact parameter $k$, defined as
$k=\eij/\epsilon$ are listed for the 1TIT molecule.

The lengths $\lij$, in a generic $N+1$ aminoacid protein,  are defined as follows.
Let us represent
the aminoacid $i$ with its
$N_{i}-C_{\alpha,i}-C_{i}$ sequence. 
Taking the native state as the reference configuration, 
$\lij$ is chosen as the 
distance between the midpoint of 
the $C_{i}$ and $N_{i+1}$ atoms and the midpoint of the $C_{j}$ and $N_{j+1}$  atoms, see figure \ref{lij_fig}.  
If $j = i+1$, $l_{i,i+i}$ corresponds to the length of  a single aminoacid.
If $i=0$ the first point is substituted with the position
of the $N_{1}$ atom, while when $j=N+1$ we take the position of  the $C_{N+1}$ atom. 

In order to fix the energy scale $\epsilon$, we define the dimensionless unfolding temperature of our model $\tT_m$, at zero force, as that temperature where the molecule order parameter $m=1/N \sum_k \average{m_k}$ is equal to 1/2.
For a given molecule, using the experimental value of the melting temperature $T_m$, defined as the temperature where half of the sample is unfolded, we define the energy scale
as $\epsilon\equiv T_m k_B/\tT_m$. This approach was used in Refs.~\cite{Ap1a,Ap3a}.

Here and in the main text we indicate the proteins either with their common name or with their pdb code.

We find that the midpoint temperature of the 1TIT model molecule, at zero force, takes the value $\tT\simeq 8.03$.
The experimentally measured unfolding temperature for such a protein is $T_m\simeq 346$~K \cite{PTPa}.
Consequently, we choose the energy scale to be $\epsilon/k_B =43.1$~K.

The midpoint temperature of the PIN1  molecule (pdb code 1I6C), at zero force, 
is found to be  $\tT\simeq 7.55$.
The experimentally measured unfolding temperature for such a protein is $T_m\simeq 332$~K \cite{MJa}.
Consequently, we choose the energy scale to be $\epsilon/k_B =44$~K.


\begin{table}
\begin{tabular} {c|c|c|c|c|c}
\begin{tabular} {|c|c|c|}
\hline
i & j & k \\
1 & 76 & 4 \\
2 & 24 & 1 \\
2 & 25 & 9 \\
2 & 26 & 11 \\
2 & 29 & 1 \\
2 & 75 & 3 \\
2 & 76 & 5 \\
2 & 77 & 1 \\
3 & 24 & 2 \\
3 & 25 & 2 \\
3 & 76 & 1 \\
3 & 77 & 3 \\
4 & 6 & 4 \\
4 & 22 & 6 \\
4 & 23 & 1 \\
4 & 24 & 3 \\
4 & 77 & 6 \\
4 & 78 & 2 \\
4 & 79 & 2 \\
5 & 23 & 4 \\
5 & 24 & 3 \\
5 & 25 & 1 \\
6 & 22 & 4 \\
6 & 23 & 6 \\
7 & 21 & 1 \\
7 & 22 & 2 \\
7 & 79 & 1 \\
8 & 20 & 5 \\
8 & 21 & 3 \\
8 & 22 & 6 \\
8 & 33 & 5 \\
8 & 70 & 7 \\
8 & 71 & 1 \\
8 & 72 & 4 \\
8 & 79 & 10 \\
8 & 80 & 4 \\
8 & 81 & 9 \\
9 & 20 & 2 \\
9 & 81 & 4 \\
9 & 82 & 1 \\
\hline
\end{tabular} &
\begin{tabular} {|c|c|c|}
\hline
i & j & k \\
10 & 20 & 1 \\
10 & 81 & 3 \\
10 & 82 & 7 \\
10 & 83 & 1 \\
11 & 18 & 4 \\
11 & 19 & 3 \\
11 & 20 & 4 \\
11 & 82 & 1 \\
11 & 83 & 9 \\
11 & 84 & 1 \\
12 & 83 & 1 \\
12 & 84 & 2 \\
12 & 86 & 1 \\
13 & 15 & 1 \\
13 & 16 & 7 \\
13 & 17 & 1 \\
13 & 18 & 1 \\
13 & 62 & 4 \\
13 & 83 & 2 \\
13 & 84 & 2 \\
13 & 85 & 4 \\
13 & 86 & 2 \\
14 & 16 & 7 \\
14 & 86 & 6 \\
15 & 61 & 2 \\
15 & 62 & 4 \\
15 & 63 & 3 \\
15 & 64 & 2 \\
15 & 85 & 6 \\
15 & 86 & 2 \\
16 & 61 & 11 \\
16 & 62 & 2 \\
17 & 60 & 2 \\
17 & 61 & 2 \\
17 & 62 & 1 \\
18 & 58 & 3 \\
18 & 59 & 2 \\
18 & 60 & 9 \\
18 & 62 & 1 \\
19 & 57 & 1 \\
\hline
\end{tabular} &
\begin{tabular} {|c|c|c|}
\hline
i & j & k \\
19 & 58 & 1 \\
19 & 59 & 7 \\
19 & 60 & 1 \\
19 & 62 & 1 \\
19 & 83 & 3 \\
20 & 56 & 1 \\
20 & 57 & 2 \\
20 & 58 & 5 \\
21 & 33 & 8 \\
21 & 55 & 1 \\
21 & 56 & 1 \\
21 & 57 & 12 \\
21 & 59 & 2 \\
21 & 70 & 10 \\
21 & 81 & 3 \\
21 & 82 & 1 \\
21 & 83 & 7 \\
22 & 33 & 2 \\
22 & 55 & 1 \\
22 & 56 & 4 \\
22 & 57 & 1 \\
23 & 33 & 3 \\
23 & 54 & 1 \\
23 & 55 & 8 \\
23 & 72 & 6 \\
23 & 79 & 6 \\
24 & 53 & 1 \\
24 & 54 & 6 \\
24 & 55 & 1 \\
25 & 29 & 1 \\
25 & 53 & 1 \\
25 & 54 & 1 \\
25 & 55 & 5 \\
25 & 72 & 2 \\
25 & 74 & 2 \\
25 & 77 & 2 \\
27 & 29 & 1 \\
28 & 52 & 1 \\
28 & 53 & 6 \\
29 & 52 & 1 \\
\hline
\end{tabular} &
\begin{tabular} {|c|c|c|}
\hline
i & j & k \\
29 & 53 & 1 \\
30 & 52 & 1 \\
30 & 55 & 2 \\
30 & 74 & 1 \\
30 & 75 & 1 \\
31 & 50 & 1 \\
31 & 55 & 1 \\
32 & 50 & 1 \\
32 & 55 & 1 \\
32 & 72 & 4 \\
32 & 73 & 2 \\
32 & 74 & 4 \\
33 & 48 & 1 \\
33 & 72 & 5 \\
33 & 73 & 3 \\
34 & 39 & 1 \\
34 & 40 & 1 \\
34 & 48 & 4 \\
34 & 55 & 3 \\
34 & 56 & 1 \\
34 & 57 & 14 \\
34 & 70 & 4 \\
34 & 71 & 1 \\
34 & 72 & 10 \\
34 & 73 & 1 \\
35 & 37 & 14 \\
35 & 38 & 10 \\
35 & 39 & 8 \\
35 & 40 & 2 \\
35 & 69 & 1 \\
35 & 70 & 1 \\
35 & 71 & 4 \\
35 & 72 & 1 \\
35 & 73 & 1 \\
36 & 38 & 4 \\
36 & 40 & 3 \\
36 & 59 & 1 \\
36 & 66 & 9 \\
36 & 69 & 1 \\
36 & 70 & 1 \\
\hline
\end{tabular} &
\begin{tabular} {|c|c|c|}
\hline
i & j & k \\
36 & 71 & 1 \\
36 & 83 & 2 \\
37 & 68 & 2 \\
37 & 69 & 4 \\
38 & 71 & 1 \\
41 & 46 & 3 \\
41 & 59 & 1 \\
42 & 46 & 1 \\
43 & 46 & 4 \\
43 & 48 & 3 \\
44 & 46 & 6 \\
46 & 59 & 1 \\
46 & 60 & 2 \\
46 & 61 & 2 \\
47 & 57 & 1 \\
47 & 59 & 2 \\
47 & 60 & 1 \\
48 & 57 & 1 \\
48 & 58 & 9 \\
48 & 59 & 1 \\
48 & 60 & 4 \\
49 & 56 & 1 \\
49 & 57 & 7 \\
49 & 58 & 1 \\
50 & 56 & 5 \\
50 & 57 & 1 \\
50 & 58 & 1 \\
51 & 54 & 2 \\
51 & 55 & 8 \\
51 & 56 & 2 \\
52 & 54 & 2 \\
52 & 55 & 1 \\
53 & 55 & 3 \\
56 & 72 & 3 \\
58 & 70 & 2 \\
60 & 62 & 2 \\
60 & 66 & 6 \\
60 & 83 & 1 \\
63 & 66 & 4 \\
63 & 85 & 5 \\
\hline
\end{tabular} &
\begin{tabular} {|c|c|c|}
\hline
i & j & k \\
64 & 66 & 5 \\
64 & 85 & 6 \\
65 & 67 & 4 \\
65 & 85 & 12 \\
65 & 86 & 2 \\
67 & 83 & 3 \\
67 & 85 & 5 \\
68 & 83 & 1 \\
68 & 84 & 3 \\
68 & 85 & 3 \\
69 & 82 & 1 \\
69 & 83 & 4 \\
69 & 84 & 1 \\
70 & 81 & 2 \\
70 & 82 & 2 \\
70 & 83 & 1 \\
71 & 79 & 1 \\
71 & 80 & 1 \\
71 & 81 & 6 \\
71 & 82 & 1 \\
71 & 83 & 3 \\
72 & 79 & 2 \\
72 & 80 & 3 \\
72 & 81 & 1 \\
73 & 78 & 2 \\
73 & 79 & 7 \\
73 & 80 & 1 \\
74 & 77 & 1 \\
74 & 78 & 4 \\
75 & 77 & 5 \\
75 & 78 & 1 \\
& &\\& &\\& &\\& &\\& &\\& &\\& &\\& &\\& &\\
\hline
\end{tabular}
\end{tabular}
\caption{Map of the interaction parameter $k$, defined as $k=\epsilon_{ij}/\epsilon$, for the 1TIT protein.}
\label{tab_app}
\end{table}


\end{document}